\documentclass[aps,pre,amsmath,amssymb,twocolumn,groupedaddress,showpacs]{revtex4}
\usepackage{graphicx}
\usepackage[utf8]{inputenc}

\usepackage{epsf}

\newcommand{\be}{\begin{equation}}
\newcommand{\ee}{\end{equation}}
\newcommand{\ba}{\begin{eqnarray}}
\newcommand{\ea}{\end{eqnarray}}
\newcommand{\baa}{\begin{eqnarray}}
\newcommand{\eaa}{\end{eqnarray}}
\newcommand{\ed}{\end{document}}

\renewcommand{\baselinestretch}{1.2}
\date{\today}

\begin{document}

\title{Bernoulli's formula and Poisson's equations  for a confined quantum gas:
\\Effects due to a moving piston}
\author{Katsuhiro Nakamura$^{(1,2)}$,  Zarifboy A. Sobirov$^{(3)}$, Davron U. Matrasulov$^{(4)}$, Sanat K.Avazbaev$^{(5)}$}
\affiliation{$^{(1)}$Faculty of Physics, National University of Uzbekistan, Vuzgorodok, Tashkent 100174, Uzbekistan\\
$^{(2)}$Department of Applied Physics, Osaka City University, Osaka 558-8585, Japan\\
$^{(3)}$Applied Mathematics Department, Tashkent Financial Institute, 60A  Amir Temur Str., Tashkent 100000, Uzbekistan\\
$^{(4)}$Turin Polytechnic University in Tashkent, 17 Niyazov str. (Small Ring), Tashkent 100094, Uzbekistan\\
$^{(5)}$ARC Centre for Antimatter-Matter Studies, Department of Applied Physics, Curtin University, GPO Box U1987, Perth 6845, Australia
}

\begin{abstract}
We study a nonequilibrium equation of states of an ideal quantum gas confined in the
cavity under a moving piston with a small but finite velocity in the case that
the cavity wall suddenly begins to move at time origin.  Confining to the
thermally-isolated process, quantum non-adiabatic (QNA) contribution to Poisson's adiabatic equations and to
Bernoulli's formula which bridges the pressure and internal energy is
elucidated. We carry out a statistical mean of the non-adiabatic (time-reversal-symmetric) force
operator found in our preceding paper (K. Nakamura et al, Phys. Rev.
{\bf E83}, 041133 (2011)) in both the low-temperature quantum-mechanical  and
high temperature quasi-classical regimes. The QNA contribution, which is
proportional to square of the piston's velocity and to inverse of the longitudinal size of the cavity, has a coefficient dependent on temperature, gas density and dimensionality of the cavity. The investigation is done for a unidirectionally-expanding 3-d
rectangular parallelepiped cavity as well as its 1-d version. Its
relevance in a realistic nano-scale heat engine is discussed.
\end{abstract}
\maketitle

\section{Introduction}
The equation of states plays an important role in thermodynamics and statistical mechanics. Let's consider the Carnot's thermodynamic cycle proposed almost two centuries ago \cite{rf:carn}. It is the most efficient cycle for converting a given heat into work. In this cycle, the system is assumed to undergo a series of different thermodynamic states and performs work on its surroundings, thereby acting as a Carnot heat engine. However, such a perfect engine is only a theoretical limit and practical engines must incorporate the effect of non-zero velocity of the moving piston.

In the Carnot cycle, the pressure ($P$) and volume ($V$) of an ideal classical gas (Boltzmann gas) confined in the cavity are assumed to obey the equilibrium equation of states, i.e., Boyle-Charles' law (BCL) and a set of Poisson's adiabatic equations in the isothermal and thermally adiabatic processes, respectively. The Poisson's adiabatic equations are derived from the first law of thermodynamics together with BCL. BCL itself is a special limit of the Bernoulli's formula (BF)  bridging between the pressure ($P$) and internal energy ($U$) for quantum and classical gas in the cavity in d-dimensions. BF is available from the relation $PV=-\Omega$ with use of density of states in calculating the thermodynamic potential $\Omega$ for both classical and quantum gas. To be specific, $PV=\frac{2}{3}U$, $U$, and $2U$ for  d=3,2, and 1, respectively. The last case may be better rewritten as $FL=2U$ with use of the force ($F$) and the length ($L$) of the 1-d cavity. For a classical gas, $U=\frac{3}{2}NkT$, $NkT$ and $\frac{1}{2}NkT$ for  d=3,2 and 1, respectively, with use of the number of particles $N$, Boltzmann constant $k$ and temperature $T$. Then the Bernoulli's formula reduces to BCL, $PV=NkT$, irrespective of dimensionality. For a quantum gas, the Bernoulli's formula works as well, where $U=E_0\left(1+0.0713 \left(mT/\hbar^2 \right)^2\left(V/N \right)^{4/3} \right)$ with $E_0=\left(3/10 \right)\left(6\pi^2 \right)^{2/3}\left(\hbar^2/m \right)\left(N/V \right)^{2/3} N  $ for d=3 Fermi gas in the low-temperature and high-density regime (see Landau-Lifshitz \cite{rf:landau}). In the thermally adiabatic process,  a set of Poisson's adiabatic equations also works, which are given by $PV^{(d+2)/d}$=const., $\frac{P}{T^{(d+2)/2}}$=const., and $VT^{d/2}$=const., irrespective of classical and quantal systems \cite{callen}.

In constructing Bernoulli's formula, the velocity of the wall of a gas container (cylinder, cavity, billiard, {\it etc}.) is assumed to be negligibly small. To make the theory of heat engines more realistic, one must evaluate the effect of the non-zero velocity of the piston, i.e., the wall motion of the gas container. Since the kinetic theory of Boltzmann gas tells that a moving piston does not play a role in the equation of states, we shall investigate the nonadiabatic dynamics in the quantum heat engine. While in recent years there appeared papers which treated
the quantum engine, they were either concerned with a quantum analog of Carnot's engine \cite{rf:bend-1,rf:bend-2,rf:abe-1,rf:abe-2} or with a quantum analog of nonequilibrium work relation (i.e., fluctuation theorem)\cite{rf:teif,rf:quan}. And no work so far was engaged in  nonadiabatic force and pressure due to a moving piston and in the statistical treatment of a noninteracting Fermi gas.

In this paper, confining ourselves to the thermally-isolated process, we shall investigate the non-equilibrium equation of states for an ideal quantum gas (Fermi gas) confined into an expanding cavity
in the case that the cavity wall suddenly begins to move at time origin. Quantum non-adiabatic (QNA) contributions  to the Bernoulli's formula and to Poisson's adiabatic equations due to the non-zero velocity of the moving piston is elucidated.  In Sec. \ref{nonad-press}, with use of the nonadiabatic force operator in our preceding paper\cite{rf:non-ad-f}, the adiabatic and non-adiabatic pressures are defined. In Sec.\ref{expect-press} expectation of nonequilibrium pressure  is expressed in terms of density of states, which will enable
the calculation of thermodynamic and statistical averages of nonadiabatic pressure. In Secs.\ref{case-1d-cav} and \ref{case-3d-cav}, the unidirectionally-expanding cavities in d=1 and d=3 dimensions are studied and explicit forms for QNA contributions to the pressure, internal energy and equation of states will be given in the low-temperature and high density regime as well as in the high-temperature and low-density regime. In Sec.\ref{phys-impli}
physical implications of QNA contributions will given. Section \ref{summary} is devoted to summary and discussions. In Appendices \ref{low-temp-myu} and \ref{high-temp-myu},
using the Fermi-Dirac distribution, we summarize several formulas for thermodynamic averages in the cases of
$1$-d and $3$-d rectangular cavities.

The isothermal process which requires a contact with the heat reservoir is outside the scope of the paper and will be investigated in due course.

\section{Adiabatic and nonadiabatic pressures}\label{nonad-press}
Before embarking upon the adiabatic and nonadiabatic pressures, we shall briefly summarize the derivation of  the adiabatic and nonadiabatic  force operators  in our preceding paper \cite{rf:non-ad-f}, but here in the context of the parallel-piped rectangular 3-d cavity.
Let's consider a Fermi gas (non-interacting Fermi particles) confined in a cavity with a moving wall (i.e., piston). The
wall receives the force from the Fermi gas
in the cavity. Under the condition that whole system consisting of Fermi particles and a moving wall keeps the energy conservation, the work done on the wall by the force is supplied by the excess energy due to the energy loss of Fermi particles
showing the non-adiabatic transition. In this way one can conceive both the adiabatic
and nonadiabatic forces. In the adiabatic limit, the adiabatic force due to the quantal gas on the cavity wall is proportional to the derivative of the confining energy with respect to the cavity size. What is a characteristic feature of the nonadiabatic force coming from the non-adiabatic transition?

\begin{figure}[htb]
\centerline{\includegraphics[width=\columnwidth]{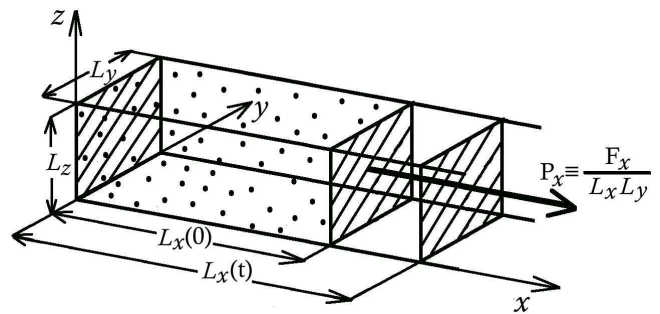}}
\caption{3-d rectangular parallelepiped cavity confining the quantal gas, with the size $L_x(t) \times L_y \times L_z$. One of its walls is moving in $x$-direction. $P_x(t)$ and $F_x(t)$ stand for the $x$-components of pressure and force.}
\label{cavity}
\end{figure}

We choose a 3-d rectangular parallelepiped cavity with the size $L_x \times L_y \times L_z$, one of whose walls is moving in $x$-direction (see Fig. \ref{cavity}).

The original Hamiltonian  for the  cavity with a time-dependent longitudinal size $L_x(t)$ is given by
\begin{align}
H_{total}=H+H_{\perp}
\label{tot-H}
\end{align}
with
\begin{eqnarray}\label{orig-H}
H&=&-\frac{\hbar^2}{2m}\frac{\partial^2}{\partial x^2},\nonumber\\
H_{\perp}&=&-\frac{\hbar^2}{2m}(\frac{\partial^2}{\partial y^2}+\frac{\partial^2}{\partial z^2}).
\end{eqnarray}
The wavefunction is a product of the longitudinal and perpendicular parts:
\begin{align}
\psi_{total}(x,y,z,t)=\psi(x,t)  \psi_{\perp}(y,z,t),
\label{orig-wavefunc}
\end{align}
which satisfies
the moving and static Dirichlet boundary conditions for $\psi$ and $\psi_{\perp}$, respectively as
\begin{align}
\label{dirch}
\psi(x=0,t)=\psi(x=L_x(t),t)=0,\\
\psi_{\perp}(y=0,z,t)=\psi_{\perp}(y=L_y,z,t)=0,\\
\psi_{\perp}(y,z=0,t)=\psi_{\perp}(y,z=L_z,t)=0.
\end{align}
Throughout the time evolution, the instantaneous (adiabatic) eigenstate is characterized
by a set of quantum numbers $(n_x,n_y,n_z)$. The longitudinal perturbation in $H$ commutes with
the perpendicular part $H_{\perp}$ in the total Hamiltonian in Eq.(\ref{tot-H}), and thereby the quantum numbers
$n_y$, and $n_z$ are conserved against an expansion along $x$.
Therefore, if a confined particle is initially in a manifold with the fixed $n_y$ and $n_z$ and the cavity expands only in $x$-direction, there occurs no mixing among manifolds with different $n_y$ and $n_z$. Consequently
the dynamics of $\psi_{total}(x,y,z,t)$  is determined by  the time-dependent Schr\"odinger equation for the longitudinal part $\psi(x,t)$ as
\begin{equation}
i\hbar \frac{\partial }{\partial t}\psi(x,t) =H\psi(x,t).
\label{orig-Schr}
\end{equation}

The expectation of the normal component of the force acting on the wall is obtained by
\begin{equation}
\bar{F}_x=-\frac{\partial}{\partial L_x(t)} \langle \psi |H|\psi \rangle
=- \langle \psi |\frac{\partial H}{\partial L_x(t)}|\psi \rangle,
\label{averF}
\end{equation}
where, in obtaining the last expression, we used  $\frac{\partial}{\partial L_x} |\psi \rangle=\frac{1}{\dot{L}_x} \frac{\partial}{\partial t} |\psi \rangle=\frac{1}{i\hbar \dot{L}_x}H|\psi \rangle$ and its Hermitian conjugate.
Hence the force operator is defined by
\begin{equation}
\hat{F}_x= -\frac{\partial H}{\partial L_x(t)}.
\label{opeF}
\end{equation}
Since the original Hamiltonian $H$ for the cavity with its time-dependent longitudinal size $L_x(t)$  does not formally include
$L_x(t)$ explicitly, however,  there is no way to define the force operator directly by using Eq.(\ref{opeF}).

To overcome this difficulty, we shall make the time-dependent canonical transformation of  $H$ related to the scale transformation of both the coordinate $x$ and
amplitude of the wave function $\psi$. This transformation  is defined by\cite{rf:TUT}
\begin{align}\label{cano-tran}
\tilde{H}=e^{-iU}(H-i\hbar \frac{\partial}{\partial t} )e^{iU},
\end{align}
with $U=-\frac{1}{2\hbar}(\hat{x}\hat{p}+\hat{p}\hat{x})\ln
L_x(t)=i\left(x\frac{\partial}{\partial x}+\frac{1}{2} \right)\ln L_x(t)$.
This canonical transformation  leads to the scaled coordinate $\tilde{x}$
defined by $e^{-iU} x e^{iU}=\tilde{x}L_x(t)$ and the scaled wave function
$\tilde{\phi}(\tilde{x},t)= e^{-iU}\psi(x,t)=\sqrt{L_x}\psi(\tilde{x}L_x,t)$.
The range of $\tilde{x}$ is $0 \le \tilde{x} \le 1$, which is time-independent.
Also the normalization factor of $\tilde{\phi}(\tilde{x},t)$ becomes $L_x$-independent
and satisfies the fixed Dirichlet boundary condition $\tilde{\phi}(0,t)=\tilde{\phi}(1,t)=0$.

Finally the Schr\"odinger equation is transformed to
\begin{align}
i\hbar \frac{\partial \tilde{\phi}}{\partial t}=\tilde{H}\tilde{\phi}
\label{scaled_Schr}
\end{align}
with the new Hamiltonian
\begin{align}
\tilde{H}= -\frac{\hbar^2}{2mL_x^2}\frac{\partial^2 }{\partial \tilde{x}^2}+
i\hbar\frac{\dot{L}_x}{L_x}\tilde{x}\frac{\partial }{\partial \tilde{x}}+\frac{i\hbar}{2} \frac{\dot{L}_x}{L_x},
\label{scaled_Hamil}
\end{align}
which is Hermitian.
$\tilde{\phi}(x,t)$  now satisfies the fixed Dirichlet boundary condition $\tilde{\phi}(0,t)=\tilde{\phi}(1,t)=0$.

Taking $L_x$ derivative of $\tilde{H}$, we can rigorously define the force operator in the transformed space, whose  inverse canonical transformation gives the force operator expressed in the original space as

\begin{align}
\hat{F}=\frac{\hat{p}^2}{mL}-\frac{\dot{L}}{2L^2}(\hat{x}\hat{p}+\hat{p}\hat{x})
=-\frac{\hbar^2}{m}\frac{1}{L} \partial_x^2+i\frac{\hbar}{2}\frac{\dot{L}}{L^2}(x\partial_x+\partial_x x),
\label{force-x}
\end{align}
where we suppressed the suffix $x$ in both the force operator and the longitudinal length.
The issue in Eq.(\ref{force-x}) is universal, irrespective of the kind of canonical transformations. In fact,
one may choose another canonical transformation such as a combination of $U$ in Eqs.(\ref{cano-tran}) and  the gauge transformation (see \cite{rf:non-ad-f}), which also guarantees the wave function to satisfy the fixed Dirichlet boundary condition and the transformed Hamiltonian, say $\tilde{H}'$,
to be Hermitian.    The derivative of  $\tilde{H}'$ w.r.t. $L_x$ defines the force operator $\hat{F}'$, and the inverse of a combination of the gauge and scale transformations results in the identical expression for $\hat{F}$.

In the final expression of Eq.(\ref{force-x}), the first and the second parts  defines the adiabatic and nonadiabatic forces, respectively.
The latter part, which gives an essential contribution when the system is not in the instantaneous eigenstates,
is invariant under the time-reversal operation since both $\dot{L}$ and $\hat{p}$ change their signs.
The expression in Eq.(\ref{force-x}) is the force normal to the wall, and, when divided by an area of the wall, it gives the
adiabatic and nonadiabatic pressures ($\hat{P}$) acting on the moving wall of the 3-d rectangular parallelepiped cavity:
\begin{align}\label{3d-pres-op}
\hat{P}=\frac{\hat{F}}{L_yL_z}.
\end{align}

\section{Expectation of nonequilibrium pressure in terms of density of states}\label{expect-press}
\begin{figure}[htb]
\centerline{\includegraphics[width=\columnwidth]{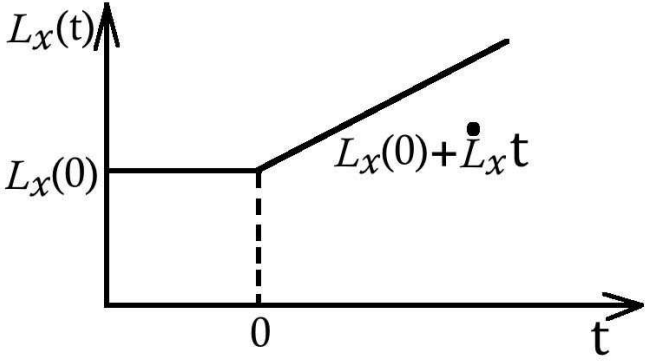}}
\caption{Time dependence of $L_x(t)$. The wall is assumed to begin to move at time origin.}
\label{mov-wall}
\end{figure}

Let' s consider the system to be thermally isolated and the wall of the cavity to begin to move
at $t=0$ suddenly (see Fig. \ref{mov-wall}). The Fermi gas in the cavity is assumed to satisfy the equilibrium Fermi-Dirac distribution until $t=0$.
The expectation of the force operator is evaluated in terms of the density operator $\rho$:
\begin{align}
\bar{F}=\rm{Tr}(\rho\hat{F}).
\end{align}

The density operator
$\rho$ for a thermally-isolated nonequilibrium state of the Fermi gas obeys von Neumann equation
\begin{align}
i\hbar \frac{\partial \rho}{\partial  t}=\left[H,\rho\right]
\end{align}
where the original Hamiltonian $H$ and coordinate $x$ are used.

Using adiabatic bases (instantaneous eigenstates) $\{|n \rangle \}$, the matrix elements of $\rho$
satisfies
\begin{align}
\dot{\rho}_{nm}=\frac{1}{i\hbar}(E_n-E_m)\rho_{nm}-\frac{\dot{L}}{L}\left(\sum_{\ell\neq n}\gamma_{\ell n}\rho_{\ell m}+\sum_{\ell\neq m}\gamma_{\ell m}\rho_{n \ell} \right)
\end{align}
with $\gamma_{mn}=(-1)^{m+n+1}\frac{2mn}{m^2-n^2}(1-\delta_{mn})$.

Then $\bar{F}$ becomes
\begin{align}
\bar{F}=\sum_{m,n}\rho_{nm} F_{mn},
\end{align}
where
\begin{align}
F_{mn}=\frac{\hbar^2}{m} \frac{(n\pi)^2}{L^3}\delta_{mn}+\frac{i\hbar\dot{L} }{L^2}\gamma_{mn}.
\end{align}

To make the problem tractable,
we assume $\dot{L}\ll v_F$, that is, the wall velocity $\dot{L}$ is much less than the Fermi velocity $v_F$, which guarantees a confined particle to collide with the cavity wall many times during the wall displacement of $O(L)$. The above unequality is scaled by $L$ and is written as
\begin{align}\label{small-vel}
\frac{\dot{L}}{L}\ll\frac{1}{\tau_F}
\end{align}
where $\tau_F(=\frac{L}{v_F})$ is a characteristic time for a particle to travel through the cavity.
With use of the smallness parameter $\frac{\dot{L}}{L}$,
we substitute the expansion
\begin{align}\label{rho-expa}
\rho=f(H)+\frac{\dot{L}}{L} g_1+ \left(\frac{\dot{L}}{L}  \right)^2g_2
\end{align}
into the von Neumann equation.
Then, for orders of $(\frac{\dot{L}}{L})^0$, $(\frac{\dot{L}}{L})^1$ and $(\frac{\dot{L}}{L})^2$,
we have
$\dot{f}(H)=0$,
$\dot{g}_{1nm}=\frac{E_n-E_m}{i\hbar} g_{1nm}-(\gamma_{mn}f_m+\gamma_{nm}f_n)$,
and
$\dot{g}_{2nn}=- \sum_\ell \gamma_{\ell n}(g_{1\ell n}+g_{1n\ell})$, respectively.

The condition Eq.(\ref{small-vel}) guarantees $\dot{L}t \ll L$
in a wide time range. Then a set of the above equations can be solved as,

\begin{align}\label{ini-FD}
f_{nm}= \frac{1}{e^{\beta(E_n-\mu)}+1} \delta_{nm}\equiv f_n \delta_{nm},
\end{align}
\begin{align}
g_{1nm}= \frac{i\hbar\gamma_{mn}}{E_n-E_m}  \left(1-e^{\frac{E_n-E_m}{i\hbar}}t \right)(f_n-f_m),
\end{align}
\begin{eqnarray}
g_{2nn}&=&-2\sum_{\ell\neq n } \gamma_{n \ell}^2 (f_n-f_\ell) \left(\frac{\hbar}{E_n-E_\ell} \right)^2 \nonumber\\
&\times&\left(1-\cos\left(\frac{E_n-E_\ell}{\hbar}t \right) \right).
\end{eqnarray}
Equation (\ref{ini-FD}) denotes the initial Fermi-Dirac distribution with inverse temperature $\beta=\frac{1}{kT}$.

Expectation value $\bar{F}$ is given by
\begin{align}
\bar{F}=\bar{F}_1+\bar{F}_2+\bar{F}_3
\end{align}
where
\begin{align}
\bar{F}_1=\sum_nf_nF_{nn}=\frac{\hbar^2}{m}\sum_n\frac{(n\pi)^2}{L^3}f_n=\frac{2}{L} \sum_n E_nf_n,
\end{align}
\begin{eqnarray}
\bar{F}_2&=&\frac{\dot{L}}{L}\sum_{n\neq m}g_{1nm}F_{mn} \nonumber\\
&=&-4\hbar^2\frac{\dot{L}^2}{L^3}\sum_{m\neq n}\frac{E_nE_m}{(E_n-E_m)^2}\frac{f_n-f_m}{E_n-E_m}\nonumber\\
&\times&2\sin^2\left(\frac{E_n-E_m}{2\hbar}t\right) \nonumber\\
&=&-8\pi^2\hbar^2\frac{\dot{L}^2}{L^3}\sum_{m\neq n}E_nE_m\frac{f_n-f_m}{E_n-E_m}\delta(E_n-E_m),\nonumber\\
\end{eqnarray}
\begin{eqnarray}
\bar{F}_3&=&\frac{\hbar^2}{m}\left(\frac{\dot{L}}{L} \right)^2  \sum_n g_{2nn} F_{nn}\nonumber\\
&=& -16\frac{\dot{L}^2}{L^3} \sum_{n>m} E_nE_m \frac{f_n-f_m}{E_n-E_m}\nonumber\\
&\times&\left(\frac{\hbar}{E_n-E_m} \right)^2 \times 2\sin^2\left(\frac{E_n-E_m}{2\hbar}t\right) \nonumber\\
&=& -32\pi^2\hbar^2\frac{\dot{L}^2}{L^3} \sum_{n>m} E_nE_m \frac{f_n-f_m}{E_n-E_m}\delta(E_n-E_m).\nonumber\\
\end{eqnarray}
It should be noted that both $\bar{F}_2$ and $\bar{F}_3$ are quadratic in $\dot{L}$. The absence of $\dot{L}$-linear terms is caused by a subtle cancellation of  the linear cross-coupling terms among the matrix elements of the force operator and those of the density matrix both expressed as a series expansion w.r.t.  $\dot{L}$.

In obtaining the final expression for $\bar{F}_1$,  $\bar{F}_2$ and  $\bar{F}_3$, we used  $E_n\equiv\frac{\pi^2\hbar^2n^2}{2mL^2}$ and the asymptotic form,
\begin{align}\label{asympt}
\frac{\sin\left(\Delta E\frac{t}{\hbar} \right)}{\Delta E} \approx \pi\delta(\Delta E),
\end{align}
which is valid in the time domain much larger than the minimum resolution of time ($t \gg \frac{\hbar}{\Delta E}$).
Thanks to Eq.(\ref{asympt}), the explicit time dependence of  $\bar{F}_2$ and $\bar{F}_3$ are suppressed.

The discrete summations can now be reduced to continuum integrations with use of 1-d density of states as
\begin{align}
\sum_{n=1}^\infty=\sum_{k_n\left(\equiv\frac{\pi n}{L} \right)}= \int_{E_0}^\infty D_1(E) dE.
\end{align}
Noting $E=\frac{\hbar^2k^2}{2m}$, $D_1(E)$ is given by
\begin{align}\label{1d-dos}
D_1(E)= \frac{dk/(\pi/L)}{dE} =\frac{\sqrt{m}L}{\sqrt{2}\pi \hbar }E^{-1/2}.
\end{align}
Using the above facts, we shall write the final results for $\bar{F}_1$,  $\bar{F}_2$ and  $\bar{F}_3$:
\begin{align}\label{1d-ff1}
\bar{F}_1=\frac{2}{L} \int_0^\infty ED_1(E)f(E),
\end{align}
\begin{eqnarray}\label{1d-ff2}
\bar{F}_2&=&-8\pi^2\hbar^2\frac{\dot{L}^2}{L^3}\int_0^\infty dE \int_0^\infty dE' E E'\left.\frac{df}{dE}\right|_{E=E'} \nonumber\\
&\times& D_1(E) D_1(E') \delta(E-E'),
\end{eqnarray}
\begin{eqnarray}\label{1d-ff3}
\bar{F}_3&=&-32\pi^2\hbar^2\frac{\dot{L}^2}{L^3}\int_0^\infty dE \int_0^\infty dE' E E'\left. \frac{df}{dE}\right|_{E=E'} \nonumber\\
&\times& D_1(E) D_1(E') \delta(E-E').
\end{eqnarray}

The purpose of the present paper is to generalize Bernoulli' s formula bridging pressure and internal energy to the case of the expanding cavity.
Therefore, one should also provide general formulas for the internal energy in the case of a moving piston.
With use of the expansion in Eq.(\ref{rho-expa}) and the matrix elements
\begin{align}
(\hat{H})_{nm}=\frac{\hbar^2\pi^2n^2}{2m L^2}\delta_{nm}\equiv E_n\delta_{nm},
\end{align}
we have the internal energy $\bar{U}$,
\begin{align}
\bar{U}= {\rm{Tr}}(\rho \hat{H})=\bar{U}_1+\bar{U}_2+\bar{U}_3.
\end{align}
Here
\begin{align}
\bar{U}_1= {\rm{Tr}}(f \hat{H})=\sum_{n=1}^\infty E_n f_n=\int_0^\infty E D_1(E)f(E)dE
\end{align}
and
\begin{eqnarray}
\bar{U}_3&=&\left(\frac{\dot{L}}{L}\right)^2  {\rm{Tr}}(g_2 \hat{H})\nonumber\\
&=&-8\pi^2\hbar^2 \left(\frac{\dot{L}}{L}\right)^2 \sum_{n>m}\frac{E_nE_m}{E_n-E_m }(f_n-f_m)
\nonumber\\
&=&-8\pi^2\hbar^2 \left(\frac{\dot{L}}{L}\right)^2\int_0^\infty dE \int_0^E dE' E E'\left.\frac{df}{dE} \right|_{E=E'} \nonumber\\
&\times& D_1(E) D_1(E') \delta(E-E').
\end{eqnarray}
Noting the absence of the diagonal elements of $g_1$,
\begin{align}
\bar{U}_2=\frac{\dot{L}}{L} {\rm{Tr}}(g_1 \hat{H})=0,
\end{align}
namely, always vanishing.

\section{Case of expanding 1-d cavity}\label{case-1d-cav}
Firstly, concentrating on the expanding 1-d cavity, we shall evaluate the final expressions in the previous section in two limiting cases, i.e.,
in the low-temperature and high-density region for a degenerate quantum gas and in the high-temperture and low-density region for a quasi-classical gas.

\subsection{Low-temperature and high-density region}
Having recourse to formulas in Eqs.(\ref{1d-lt-fd0}) and (\ref{1d-lt-fd1}), the expectation of force terms in Eqs.(\ref{1d-ff1})-(\ref{1d-ff3}) becomes
\begin{eqnarray}\label{1d-f1-lt}
\bar{F}_1 &=&\frac{2\sqrt{m}}{\sqrt{2}\pi \hbar}\int_0^\infty \sqrt{E} f(E) dE\nonumber\\
&=&\frac{2\sqrt{2m}}{3\pi \hbar} \mu^{3/2}\left[1+\frac{\pi^2}{8}(kT)^2\mu^{-2}+\frac{7\pi^4}{640}(kT)^4\mu^{-4}+\cdots   \right],\nonumber\\
\end{eqnarray}
\begin{align}\label{1d-f2-lt}
\bar{F}_2=-4m \frac{\dot{L}^2}{L} \int_0^\infty dE E \frac{df}{dE}=4m\frac{\dot{L}^2}{L}\mu,
\end{align}
\begin{align}\label{1d-f3-lt}
\bar{F}_3=-8m \frac{\dot{L}^2}{L} \int_0^\infty dE E \frac{df}{dE}=8m\frac{\dot{L}^2}{L}\mu.
\end{align}
where $\mu$ is the chemical potential.
Thereby,
 \begin{eqnarray}
\bar{F}&=&\bar{F}_1+\bar{F}_2+\bar{F}_3=\frac{2\sqrt{2m}}{3\pi\hbar}\mu^{3/2}\left[1+\frac{\pi^2}{8}(kT)^2\mu^{-2}+\cdots  \right] \nonumber\\
&+&12m\frac{\dot{L}^2}{L}\mu .
\end{eqnarray}
Noting the low-temperature and high-density expansion of $\mu$ with use of  particle number $N$ in Eq.(\ref{1d-lt-che}), we have
\begin{eqnarray}\label{1d-lt-f}
\bar{F}&=&\frac{\pi^2\hbar^2}{3m} \left(\frac{N}{L} \right)^3\left(1+\frac{1}{\pi^2}\left(\frac{mkT}{\hbar^2} \right)^2\left(\frac{N}{L}\right)^{-4}+\cdots  \right)\nonumber\\
&+&6\pi^2\hbar^2\frac{\dot{L}^2}{L}\left(\frac{N}{L} \right)^2\left(1+\frac{1}{3\pi^2}\left(\frac{mkT}{\hbar^2} \right)^2\left(\frac{N}{L}\right)^{-4}+\cdots  \right).
\nonumber\\
\end{eqnarray}
Equation (\ref{1d-lt-f}) can be rewritten as
\begin{eqnarray}\label{1d-lt-po}
\bar{F}L^3&-&\frac{\pi^2\hbar^2}{3m} N^3\left(1+ \frac{\pi^2}{4}\left(\frac{kT}{\mu}\right)^2+ \cdots  \right)\nonumber\\
&=&6\pi^2\hbar^2\dot{L}^2N^2\left(1+ \frac{\pi^2}{12}\left(\frac{kT}{\mu}\right)^2+ \cdots  \right).
\end{eqnarray}
Noting $\frac{kT}{\mu}$=const. in the unperturbed adiabatic state, Eq.(\ref{1d-lt-po}) is nothing but a generalization of Poisson's adiabatic equation (PAE) in 1 dimension
in the case that a piston has a small but non-zero velocity.
A qualitatively new correction term on the right-hand side is quadratic in both velocity of the piston and particle number.

The internal energy for the expanding 1-d cavity is calculated in a similar way:
Noting
\begin{align}
\bar{U}_1=\frac{\sqrt{2m}L}{3\pi\hbar}\mu^{3/2}\left(1+\frac{\pi^2}{8}(kT)^2\mu^{-2}+\cdots  \right),
\end{align}
and
\begin{align}
\bar{U}_3=2m  \dot{L}^2\mu,
\end{align}
we have
\begin{eqnarray}
\bar{U}&=&\bar{U}_1+ \bar{U}_3\nonumber\\
&=&\frac{\sqrt{2m}L}{3\pi\hbar}\mu^{3/2}\left(1+\frac{\pi^2}{8}(kT)^2\mu^{-2}+\cdots  \right)+ 2m   \dot{L}^2\mu.\nonumber\\
\end{eqnarray}

Using the expansion for $\mu$ in Eq.(\ref{1d-lt-che}), we have
\begin{eqnarray}\label{1d-lt-int}
\bar{U}&=&\frac{\pi^2\hbar^2}{6m} \left(\frac{N}{L} \right)^2N\left(1+\frac{1}{\pi^2}\left(\frac{mkT}{\hbar^2} \right)^2\left(\frac{N}{L}\right)^{-4}+\cdots  \right)\nonumber\\
&+&\pi^2\hbar^2\dot{L}^2 \left(\frac{N}{L} \right)^2 \left(1+\frac{1}{3\pi^2} \left( \frac{mkT}{\hbar^2} \right)^2\left(\frac{N}{L}\right)^{-4}+\cdots  \right).
\nonumber\\
\end{eqnarray}
The first term corresponds to the 1-d version of the existing result (Landau-Lifshitz\cite{rf:landau}), and the second one is a nonequilibrium correction.
Combining Eqs. (\ref{1d-lt-f}) and (\ref{1d-lt-int}), we have
\begin{eqnarray}
\bar{F}L&-&2\bar{U}\nonumber\\
&=&4\pi^2\hbar^2\dot{L}^2 \left(\frac{N}{L} \right)^2 \left(1+\frac{1}{3\pi^2} \left( \frac{mkT}{\hbar^2} \right)^2\left(\frac{N}{L}\right)^{-4}+\cdots  \right),
\nonumber\\
\end{eqnarray}
which generalize the Bernoulli's formula in 1-dimension. The right-hand side gives a qualitatively new correction term due to a moving piston.
This equation stands for the nonequilibrium equation of states for a quantal gas confined in the expanding cavity with the finite velocity $(\dot{L})$ of a piston.

\subsection{High-temperature and low-density region}
In this subsection we shall investigate the opposite limit, i.e., the high-temperature and low-density quasi-classical regime.

Here we shall have recourse to a high-temperature expansion of Fermi-Dirac distribution expansion
with a negative value $\mu$,
\begin{align}\label{fd-hy-expa}
f(E)\equiv \frac{1}{e^{\beta(E-\mu)}+1}=\sum_{n=1}^\infty (-1)^{n-1} e^{-n\beta (E-\mu)}.
\end{align}
Substituting Eq.(\ref{fd-hy-expa}) into middle terms in each of Eqs.(\ref{1d-f1-lt})-(\ref{1d-f3-lt}),
one can evaluate the force:
\begin{eqnarray}\label{1d-f1-ht}
\bar{F}_1&=&\frac{2\sqrt{m}}{\sqrt{2}\pi\hbar}\sum_{n=1}^\infty (-1)^{n-1}\int_0^\infty E^{1/2} e^{-n\beta (E-\mu)}dE
\nonumber\\
&=&\left(\sqrt{\frac{m}{2\pi\hbar^2}} \right)(kT)^{3/2}e^{\beta\mu}\left(1-\frac{e^{\beta\mu}}{2\sqrt{2}}+O(e^{2\beta\mu})  \right) ,\nonumber\\
\end{eqnarray}
\begin{eqnarray}\label{1d-f23-ht}
\bar{F}_2+\bar{F}_3&=&-12m\frac{\dot{L}^2}{L}\sum_{n=1}^\infty(-1)^n n\beta  \int_0^\infty dE E e^{-n\beta (E-\mu)}\nonumber\\
&=&12m\frac{\dot{L}^2}{L}kTe^{\beta\mu}\left(1-\frac{1}{2}e^{\beta\mu}+O(e^{2\beta\mu})  \right).
\end{eqnarray}
Therefore
\begin{eqnarray}\label{1d-f123-ht}
\bar{F}&=&\bar{F}_1+\bar{F}_2+\bar{F}_3\nonumber\\
&=&\left(\sqrt{\frac{m}{2\pi\hbar^2}} \right)(kT)^{3/2}e^{\beta\mu}\left(1-\frac{e^{\beta\mu}}{2\sqrt{2}}+O(e^{2\beta\mu})  \right)\nonumber\\
&+&12m\frac{\dot{L}^2}{L}kTe^{\beta\mu}\left(1-\frac{1}{2}e^{\beta\mu}+O(e^{2\beta\mu})  \right).
\end{eqnarray}
Equation (\ref{1d-f123-ht}) can be rewritten as
\begin{eqnarray}\label{1d-f123-po}
\frac{\bar{F}}{(kT)^{3/2}}
&-&\left(\sqrt{\frac{m}{2\pi\hbar^2}} \right)e^{\beta\mu}\left(1-\cdots  \right)\nonumber\\
&=&12m\frac{\dot{L}^2}{L}(kT)^{-1/2}e^{\beta\mu}\left(1-\cdots  \right).
\end{eqnarray}
Since $\beta \mu=$const. in the unperturbed adiabatic state (see Landau-Lifshitz \cite{rf:landau}), Eq.(\ref{1d-f123-po}) is a generalization the Poisson equation  in 1 dimension expressed in terms of
pressure and temperature.

Using in Eq.(\ref{1d-f123-ht}) a high-temperature and low-density expansion of $e^{\beta\mu}$ in Eq.(\ref{1d-ht-che}),
we have
\begin{eqnarray}
\bar{F}&=&\frac{N}{L} kT \left( 1-\frac{3\sqrt{\pi}}{2} \frac{  \frac{N}{L} \hbar} {\sqrt{mkT}}
+O\left( \frac{  \left(\frac{N}{L}\hbar\right)^2} {mkT}  \right)  \right)\nonumber\\
&+&12\sqrt{2\pi} \hbar \frac{\dot{L}^2}{L}\frac{N}{L}\sqrt{mkT} \left(1-\left(1+\frac{1}{\sqrt{2}}\right)\sqrt{\pi} \frac{\frac{N}{L}\hbar }{\sqrt{mkT}}    \right).
\nonumber\\
\end{eqnarray}
Similarly, the internal energy is now given by
\begin{eqnarray}
\bar{U}&=&\bar{U}_1+\bar{U}_3\nonumber\\
&=&\frac{N}{2} kT \left( 1-\frac{3\sqrt{\pi}}{2} \frac{  \frac{N}{L} \hbar} {\sqrt{mkT}}
+O\left( \frac{  \left(\frac{N}{L}\hbar\right)^2} {mkT}  \right)  \right)\nonumber\\
&+&2\sqrt{2\pi} \hbar \dot{L}^2\frac{N}{L}\sqrt{mkT} \left(1-\left(1+\frac{1}{\sqrt{2}}\right)\sqrt{\pi} \frac{\frac{N}{L}\hbar }{\sqrt{mkT}}    \right).
\nonumber\\
\end{eqnarray}

Therefore, a generalized Bernoulli's formula in the quasi-classical region is given by
\begin{align}\label{1d-bf-ht}
\bar{F}L-2\bar{U}=
8\sqrt{2\pi} \hbar \dot{L}^2\frac{N}{L}\sqrt{mkT} \left(1-\left(1+\frac{1}{\sqrt{2}}\right)\sqrt{\pi} \frac{\frac{N}{L}\hbar }{\sqrt{mkT}}   \right).
\end{align}
The right-hand side is a nonequilibrium contribution due to the finite velocity of a piston. We find
that a deviation from Bernoulli's formula appears only when the quantum effect will be incorporated.
In fact, in the limit $\hbar\to 0$, we see $\bar{U}=\frac{N}{2}kT$ and Eq.(\ref{1d-bf-ht}) becomes the 1-d version of Boyle-Charles' law,  $\bar{F}L=NkT$, which includes no contribution due to kinematics of the piston.

\section{Case of 3-d rectangular parallelepiped cavity showing a unidirectional expansion}\label{case-3d-cav}
The realistic heat engine is composed of a $3$-d cavity with a piston moving in a fixed ($x$) direction. The force $\bar{F}$ in the previous sections is taken as $x$ component of the force vector for the case of $1$-d motion of the piston in the $3$-d rectangular parallelepiped cavity with size $L_x\times L_y\times L_z$ under the fixed perpendicular (or transverse) modes ($n_y, n_z$).

We shall denote $\bar{F_x}$ as the $x$-component of the force vector averaged over both longitudinal and perpendicular modes. Noting Eq.(\ref{3d-pres-op}), the expectation of pressure on the wall of a piston is given by
\begin{align}
\bar{P}=\frac{\bar{F}_x}{L_y L_z},
\end{align}
where $L_y L_z$ is an area of the wall.

$\bar{F_x}$  can be evaluated in a similar way as $\bar{F}$, but Fermi-Dirac distribution should include a contribution of the energy due to the perpendicular modes. Namely, the eigen-energy of a particle is now
\begin{align}
E(n_x,n_y,n_z)=E_\parallel(n_x) +E_\bot (n_y,n_z)
\end{align}
with
\begin{align}
E_\parallel(n_x)=\frac{\hbar^2}{2m}\left(\frac{n_x\pi}{L_x} \right)^2,
\end{align}
\begin{align}
E_\bot(n_y, n_z)=\frac{\hbar^2}{2m}\left( \left(\frac{n_y\pi}{L_y} \right)^2+ \left(\frac{n_z\pi}{L_z} \right)^2\right),
\end{align}
and Fermi-Dirac distribution is given by
\begin{align}
f(E)=\frac{1}{e^{\beta\left[(E_\parallel+E_\bot)-\mu \right]}+1}.
\end{align}
The statistical average is the one over the longitudinal mode ($n_x$), followed by another one over the perpendicular modes $(n_y,n_z)$.
The expectation value $\bar{F_x}$ is given by
\begin{align}
\bar{F}_x=\bar{F}_{x1}+\bar{F}_{x2}+\bar{F}_{x3}.
\end{align}

In the low-temperature and high-density regime, we have the following results:
\begin{eqnarray}\label{3d-fx-1}
& &\bar{F}_{x1}=\frac{2}{L_x}\int_0^\infty dE \int_0^E dE_\parallel E_\parallel D_1(E_\parallel) D_2(E-E_\parallel) f(E)\nonumber\\
&=&\frac{8\sqrt{2}}{15\pi^2}\left(\frac{m}{\hbar^2} \right)^{3/2}L_y L_z \mu^{5/2}\left(1+\frac{5\pi^2}{8}(kT)^2\mu^{-2} \right),
\end{eqnarray}
and
\begin{eqnarray}\label{3d-fx-2}
& &\bar{F}_{x2}\nonumber\\
&=&-8\pi^2\hbar^2\frac{\dot{L}_x^2}{L_x^3}\int_0^\infty dE \int_0^E dE_\parallel \int_0^E dE_\parallel' E_\parallel E_\parallel' \left.\frac{df}{dE}\right|_{E=E_\parallel ' +E_\bot}  \nonumber\\
&\times& D_1(E_\parallel)D_1(E_\parallel')D_2(E-E_\parallel)\delta(E_\parallel'-E_\parallel)\nonumber\\
&=&\frac{4\hbar^2}{\pi}\left(\frac{m}{\hbar^2} \right)^2\frac{\dot{L}_x^2}{L_x}L_yL_z\mu^2\left(1+\frac{\pi^2}{3}(kT)^2\mu^{-2} \right),
\end{eqnarray}
where we employed the $2$-d density of states
\begin{align}
D_2(E)=\frac{2L_yL_z}{\pi}\frac{m}{\hbar^2}
\end{align}
together with $D_1(E)$ in Eq.(\ref{1d-dos}).
$\bar{F}_{x3}$ can be obtained in a similar way, but $8\pi^2\hbar^2\int_0^EdE_\parallel'$ in the integral of $\bar{F}_{x2}$ is to be replaced by $32\pi^2\hbar^2\int_0^{E_\parallel}dE_\parallel'$, which eventually leads to $\bar{F}_{x3}=2\bar{F}_{x2}$.

Then the pressure on the wall is
\begin{eqnarray}\label{3d-lt-pre}
\bar{P}&=&\frac{\bar{F}_{x1}+\bar{F}_{x2}+\bar{F}_{x3}}{L_yL_z}\nonumber\\
&=&\frac{8\sqrt{2}}{15\pi^2}\left(\frac{m}{\hbar^2} \right)^{3/2} \mu^{5/2}\left(1+\frac{5\pi^2}{8}(kT)^2\mu^{-2} \right)\nonumber\\
&+& \frac{12\hbar^2}{\pi}\left(\frac{m}{\hbar^2} \right)^2\frac{\dot{L}_x^2}{L_x}\mu^2\left(1+\frac{\pi^2}{3}(kT)^2\mu^{-2} \right).
\end{eqnarray}
With use of the low-temperature expansion of $\mu$ in Eq.(\ref{3d-lt-che}), Eq.(\ref{3d-lt-pre}) can be written as

\begin{eqnarray}\label{3d-lt-po}
\bar{P}V^{5/3}&-&\frac{3^{2/3}\pi^{4/3}}{5\times2^{2/3}}\frac{\hbar^2}{m}N^{5/3}\left(1+\frac{5\pi^2}{12}\left(\frac{kT}{\mu}\right)^2 \right)\nonumber\\
&=& \frac{3^{7/3}\pi^{5/3}\hbar^2}{2^{4/3}}N^{4/3}V^{1/3}\frac{\dot{L}_x^2}{L_x} \left(1+\frac{\pi^2}{6}\left(\frac{kT}{\mu}\right)^2 \right).
\nonumber\\
\end{eqnarray}
This is a 3-d version of Poisson's adiabatic equation which now incorporates the non-adiabtic contribution.
As in the case of the 1-d cavity the qualitatively new contribution is proportional to the square of the wall velocity and to inverse of the longitudinal size of the cavity, but the coefficient shows a different dependence on particle number.

The internal energy for the $3$-d rectangular cavity with a moving piston is straightforward:
\begin{align}
\bar{U}^{\rm 3-d}=\bar{U}_1^{\rm 3-d}+\bar{U}_3^{\rm 3-d}
\end{align}
with
\begin{align}\label{3d-u-1}
\bar{U}_1^{\rm 3-d}=\int_0^\infty dE \int_0^E dE_\parallel E D_1(E_\parallel) D_2(E-E_\parallel) f(E),
\end{align}

\begin{eqnarray}\label{3d-u-3}
\bar{U}_3^{\rm 3-d}&=&-8\pi^2\hbar^2\left(\frac{\dot{L}_x}{L_x}\right)^2\int_0^\infty dE  \nonumber\\
&\times&\int_0^E dE_\parallel \int_0^{E_{\parallel}} dE_\parallel' E_\parallel E_\parallel' \left.\frac{df}{dE}\right|_{E=E_\parallel ' +E_\bot}  \nonumber\\
&\times&
D_1(E_\parallel)D_1(E_\parallel')D_2(E-E_\parallel)\delta(E_\parallel'-E_\parallel).
\end{eqnarray}
It should be noted that, in the calculation of $\bar{U}_1^{\rm 3-d}$, the bulk energy $E(=E_\parallel+E_\bot)$ is averaged which is a $3$-d generalization of the $1$-d energy. The final result for the internal energy is
\begin{eqnarray}\label{3d-int-low-T}
\bar{U}^{\rm 3-d}&=& \frac{4\sqrt{2}}{5\pi^2}\left(\frac{m}{\hbar^2} \right)^{3/2}L_xL_y L_z \mu^{5/2}\left(1+\frac{5\pi^2}{8}(kT)^2\mu^{-2} \right)\nonumber\\
&+&\frac{2\hbar^2}{\pi}\left(\frac{m}{\hbar^2} \right)^2\frac{\dot{L}_x^2}{L_x}L_xL_yL_z\mu^2\left(1+\frac{\pi^2}{3}(kT)^2\mu^{-2} \right).\nonumber\\
\end{eqnarray}
(With use of the expansion for $\mu$ in Eq.(\ref{3d-lt-che}), the first term on r.h.s. of Eq.(\ref{3d-int-low-T})
proves to agree with the result by Landau-Lifshitz reproduced in Introduction. The minor discrepancy of a numerical prefactor of $O(1)$ is due to our choice of anisotropic density of states in Eq.(\ref{3d-u-1}) for the
3-d rectangular parallelepiped cavity.)

The Bernoulli's formula in the present case becomes:
\begin{align}\label{3d-lt-bf}
\bar{P}V-\frac{2}{3}\bar{U}^{\rm 3-d}=\frac{32\hbar^2}{3\pi}\left(\frac{m}{\hbar^2}\right)^2\frac{\dot{L}_x^2}{L_x}V\mu^2\left(1+\frac{\pi^2}{3}(kT)^2\mu^{-2} \right)
\end{align}
with $V=L_xL_yL_z$.
With use of the low-temperature expansion of $\mu$ in Eq.(\ref{3d-lt-che}), Eq.(\ref{3d-lt-bf})
can be rewritten as
\begin{eqnarray}
\bar{P}V-\frac{2}{3}\bar{U}^{\rm 3-d}&=&c_0 \hbar^2  \left(\frac{N}{V} \right)^{4/3}\frac{\dot{L}_x^2}{L_x} V \nonumber\\
&\times&\left(1+c_1  \left(\frac{V}{N} \right)^{4/3} \left(\frac{mkT}{\hbar^2}\right)^2 \right).
\end{eqnarray}
with $c_0=\frac{2(3\sqrt{2})^{4/3}\pi^{5/3}}{3}$ and $c_1=\frac{4}{3(3\sqrt{2})^{4/3}\pi^{2/3}}$. The right-hand side is a nonadiabatic contribution to the equilibrium equation of states in 3-dimensions due to a moving piston. In the quantum adiabatic limit $\dot{L}_x=0$, the above equation reduces to the standard Bernoulli's formula for the $3$-d quantum gas.

We shall proceed to the high-temperature and low-density regime. The values of $\bar{F}_{x1}$, $\bar{F}_{x2}$,
$\bar{F}_{x3}$, $\bar{U}_1^{\rm 3-d}$ and $\bar{U}_3^{\rm 3-d}$ for the $3$-d cavity are evaluated,
by substituting into the second expressions in each of Eqs. (\ref{3d-fx-1}), (\ref{3d-fx-2}), (\ref{3d-u-1}) and (\ref{3d-u-3}) the expansion of Fermi-Dirac distribution in Eq.(\ref{fd-hy-expa}). The results are:
\begin{align}
\bar{F}_{x1}=\sqrt{2}\pi^{-3/2} \left(\frac{m}{\hbar^2} \right)^{3/2} (kT)^{5/2} L_y L_z e^{\beta\mu}\left(1-2^{-5/2}e^{\beta\mu} \right),
\end{align}
\begin{align}
\bar{F}_{x2}+\bar{F}_{x3}=\frac{24\hbar^2}{\pi}\frac{\dot{L}_x^2}{L_x}  \left(\frac{m}{\hbar^2} \right)^2 (kT)^{2} L_y L_z e^{\beta\mu}\left(1-\frac{1}{4}e^{\beta\mu} \right).
\end{align}
Then the pressure defined by
\begin{align}
\bar{P}=\frac{\bar{F}_{x1}+\bar{F}_{x2}+\bar{F}_{x3}}{L_yL_z}
\end{align}
satisfies

\begin{eqnarray}
\frac{\bar{P}}{T^{5/2}}&-&\sqrt{2}\pi^{-3/2} \left(\frac{m}{\hbar^2} \right)^{3/2} e^{\beta\mu}\left(1-2^{-5/2}e^{\beta\mu} \right)\nonumber\\
&=&\frac{24\hbar^2}{\pi}\frac{\dot{L}_x^2}{L_x}  \left(\frac{m}{\hbar^2} \right)^2 \frac{1}{\sqrt{kT}} e^{\beta\mu}
\left(1-\frac{1}{4}e^{\beta\mu} \right).
\end{eqnarray}

Similarly, we see
\begin{align}
\bar{U}_1^{\rm 3-d}=\frac{3\sqrt{2}}{2}\pi^{-3/2} \left(\frac{m}{\hbar^2} \right)^{3/2} (kT)^{5/2} V e^{\beta\mu}\left(1-2^{-5/2}e^{\beta\mu} \right),
\end{align}
\begin{align}
\bar{U}_3^{\rm 3-d}=\frac{4\hbar^2}{\pi}\frac{\dot{L}_x^2}{L_x}  \left(\frac{m}{\hbar^2} \right)^2 (kT)^{2} V e^{\beta\mu}\left(1-\frac{1}{4}e^{\beta\mu} \right).
\end{align}
leading to the internal energy, $\bar{U}^{\rm 3-d}= \bar{U}_1^{\rm 3-d}+\bar{U}_3^{\rm 3-d}$.
The Bernoulli's formula is now given by
\begin{align}
\bar{P}V-\frac{2}{3}\bar{U}^{\rm 3-d}=\frac{64\hbar^2}{3\pi}\frac{\dot{L}_x^2}{L_x}  \left(\frac{mkT}{\hbar^2} \right)^2 V e^{\beta\mu}\left(1-\frac{1}{4}e^{\beta\mu} \right).
\end{align}
Noting the high-temperature and low-density expansion of $e^{\beta\mu}$ in Eq.(\ref{3d-ht-che-exp}), we see
\begin{eqnarray}
\bar{P}V&-&\frac{2}{3}\bar{U}^{\rm 3-d}=\frac{32\sqrt{2\pi}\hbar^2}{3\pi}\frac{\dot{L}_x^2}{L_x}  \left(\frac{mkT}{\hbar^2} \right)^{1/2} N\nonumber\\
&\times&\left(1+\frac{\sqrt{2}-1}{4\sqrt{2} }\pi^{3/2}\frac{N}{V}\left(\frac{\hbar^2}{mkT} \right)^{3/2} \right).
\end{eqnarray}
We can confirm that the nonadiabatic contribution (NC) appears as a quantum effect and plays a role with decreasing the system's size ($L_x$).
In other words, NC vanishes in the classical limit ($\hbar \rightarrow 0$), which is consistent with the kinetic theory of Boltzmann gas which incorporates the effect of moving piston. The essential results obtained in this Section is summarized in Table 1.
\begin{widetext}
{\bf Table 1.} Non-adiabatic contributions to equation of states in thermally-isolated process in 3 dimensions.

\begin{center}
\begin{tabular}{|c|l|l|} \hline
Equation of states & low-temperature quantal region & high-temperature quasi-classical region\\ \hline
\hline
Poisson's adiabatic equations &
$\bar{P}V^{5/3}-\frac{3^{2/3}\pi^{4/3}}{5\times2^{2/3}}\frac{\hbar^2}{m}N^{5/3}\left(1+\frac{5\pi^2}{12}\left(\frac{kT}{\mu}\right)^2+\cdots \right)$
 &
$\frac{\bar{P}}{T^{5/2}}-\sqrt{2}\pi^{-3/2} \left(\frac{m}{\hbar^2} \right)^{3/2} e^{\beta\mu}\left(1-2^{-5/2}e^{\beta\mu} \right)$\\
 &
$= \frac{3^{7/3}\pi^{5/3}\hbar^2}{2^{4/3}}N^{4/3}V^{1/3}\frac{\dot{L}_x^2}{L_x} \left(1+\frac{\pi^2}{6}\left(\frac{kT}{\mu}\right)^2+ \cdots \right)$
&
$=\frac{24\hbar^2}{\pi}\frac{\dot{L}_x^2}{L_x}  \left(\frac{m}{\hbar^2} \right)^2 \frac{1}{\sqrt{kT}} e^{\beta\mu}
\left(1-\frac{1}{4}e^{\beta\mu}+\cdots \right)$\\
\hline
Bernoulli's formula &
$\bar{P}V-\frac{2}{3}\bar{U}^{\rm 3-d}$
 &
$\bar{P}V-\frac{2}{3}\bar{U}^{\rm 3-d}$
\\
 &
$=2^{5/3}3^{1/3}\pi^{5/3}\hbar^2  \left(\frac{N}{V} \right)^{4/3}\frac{\dot{L}_x^2}{L_x} V $
&
$=\frac{32\sqrt{2\pi}\hbar^2}{3\pi}\frac{\dot{L}_x^2}{L_x}  \left(\frac{mkT}{\hbar^2} \right)^{1/2} N$\\
 &
$\times\left(1+\frac{2^{4/3}}{3^{7/3}\pi^{2/3}}\left(\frac{V}{N} \right)^{4/3} \left(\frac{mkT}{\hbar^2}\right)^2+\cdots \right)$
&
$\times \left(1+\frac{\sqrt{2}-1}{4\sqrt{2} }\pi^{3/2}\frac{N}{V}\left(\frac{\hbar^2}{mkT} \right)^{3/2} +\cdots\right)$\\
\hline

\end{tabular}

\end{center}
\end{widetext}

\section{Physical implications of quantum nonadiabatic contributions}\label{phys-impli}
So far we have obtained the completely analytical non-adiabatic contribution to the non-equilibrium equation of states in the cases of 3-d rectangular parallelepiped cavity and its 1-d version, separately. To physically interpret the obtained results, however, it is more convenient to see the non-equilibrium equations of states for the general d-dimensional hyper-rectangular cavity which has the volume $V=L^{d-1}L_x$ and the moving wall (surface) with area $S=L^{d-1}$.  Such general derivation is also possible by using the {\bf density of states in $d$ dimensions}. After tedious and lengthy calculation (to be published elsewhere),  the Bernoulli's formulas for the d-dimensional cavity are given by
\begin{equation}
\bar{P}V-\frac{2}{d}U \sim \hbar^2 V \left(\frac{N}{V}\right)^{1+\frac{1}{d}}\frac{\dot{L_x}^2}{L_x}
\label{Bel-lowT}
\end{equation}
and
\begin{equation}
\bar{P}V-\frac{2}{d}U \sim \hbar (mkT)^{\frac{1}{2}} \left(\frac{N}{V}\right)V\frac{\dot{L_x}^2}{L_x},
\label{Bel-highT}
\end{equation}
respectively, for the low-temperature high-density and high-temperature low-density regions.  In a similar way, the corresponding Poisson's adiabatic equations are
\begin{equation}
\frac{\bar{P}V^{\frac{d+2}{d}}}{const}-1 \sim m\left(\frac{N}{V}\right)^{-\frac{1}{d}}\frac{\dot{L_x}^2}{L_x}
\label{Pois-lowT}
\end{equation}
and
\begin{equation}
\frac{\bar{P} \diagup (kT)^{\frac{d+2}{2}}}{const} -1
\sim \hbar \left(\frac{m}{kT}\right)^{\frac{1}{2}} \frac{\dot{L_x}^2}{L_x},
\label{Pois-highT}
\end{equation}
respectively, for the low-temperature high-density and high-temperature low-density regions.  The apparently-extra dimensionality of energy  ($\rm [ML^2T^{-2}]$) on the right-hand sides in all of the four equations above is traced back to our simplified replacement in Eq.(\ref{asympt}) and therefore can be suppressed. Equations (\ref{Bel-lowT})-(\ref{Pois-highT}) recover all the results for $d=1$ and $d=3$ cavities in the previous Sections. We find the important features:

1) Quantum non-adiabatic (QNA) contributions are quadratic in the wall velocity and therefore time-reversal symmetric, in marked contrast to the conventional belief \cite{fric-force} that  the nonadiabatic force is linear in the wall velocity and breaks the time-reversal symmetry;

2) QNA contributions are positive, which means that the moving wall gives rise to the apparently repulsive interaction among non-interacting Fermi particles, irrespective of the direction of the wall motion, namely for both expansion and contraction of the cavity;

3) QNA contributions are inversely proportional to the longitudinal size of the cavity and
become more and more important when the cavity size is decreased. In particular, they will play
a nontrivial role in nano-scale heat engines based on quantum dots;

4) QNA contributions play an essential role in Bernoulli's formula rather than in Poisson's equation.
In fact, the coefficients prior to $\frac{\dot{L_x}^2}{L_x}$ are increased in Eq. (\ref{Bel-lowT}) and decreased in Eq. (\ref{Pois-lowT}) as particle density $\frac{N}{V}$  is increased. Similarly, the coefficients is increased in Eq. (\ref{Bel-highT}) and decreased in Eq. (\ref{Pois-highT}) as temperature is increased.

The above 4 issues constitute a  punchline of the present paper. The Poisson's adiabatic equation and Bernoulli's formula, both of which are the basic laws of thermodynamics,  are now generalized so as to include the QNA contributions that have never been reported so far.

\section{Summary and discussions}\label{summary}
Confining ourselves to the thermally-isolated process, we study a nonequilibrium equation of states of an ideal quantum gas confined to the cavity under a moving piston with a small but finite velocity.
The cavity wall is assumed to begin to move  suddenly  at time origin.
Quantum non-adiabatic (QNA) contribution to
Bernoulli's formula which bridges the pressure and internal energy is
elucidated. Statistical means of the non-adiabatic (time-reversal symmetric) force and pressure
operator \cite{rf:non-ad-f} are carried out in both the low-temperature quantum-mechanical  and
high temperature quasi-classical regimes.
QNA contributions are quadratic in the piston's velocity and therefore time-reversal symmetric, in marked contrast to the conventional belief \cite{fric-force}, and
they are positive, which means that the moving piston gives rise to the apparently repulsive interaction among non-interacting Fermi particles, for both expansion and contraction of the cavity.
QNA contributions are inversely proportional to the longitudinal size of the cavity, and thereby play
a nontrivial role in nano-scale heat engines based on quantum dots.
The investigation is done for an expanding 3-d
rectangular parallelepiped cavity as well as its 1-d version. The nonequilibrium contributions to Poisson's adiabatic equation  are also elucidated.

In the context of a classical gas, Curzon and Ahlborn\cite{rf:cu-ah} and others\cite{rf:broe,rf:sc-se,rf:izum}  investigated a {\bf finite-time}  Carnot heat engine and obtained an interesting efficiency. However, they neither considered a quantum gas nor showed a nonequilibrium equation of states due to a moving piston.
Therefore one of the directions to extend our work may be to proceed to the same analyses as given here of the isothermal process which requires a contact of nano-scale engine with a heat reservoir.  Another direction may be the fast-forwarding of the adiabatic expansion of a cavity \cite{rf:mas-1,rf:mas-2} in the framework of von Neumann equation, to see an accelerated quantum Carnot heat engine. These subjects will be investigated in due course.

{\em Acknowledgments.} One of the authors (K. N.) is grateful to Adolfo del Campo, Takaaki Monnai, Takahiro
Sagawa and Ayumu Sugita for useful comments.

\appendix
\section{Thermodynamic averages in low-temperature region at $T\ll T_0$ (degenerate temperature)}\label{low-temp-myu}
With use of the Fermi-Dirac distribution $f(E)=\frac{1}{e^{\beta(E-\mu)}+1}$, we summarize several formulas for thermodynamic averages (see Landau-Lifshitz\cite{rf:landau}) in the cases of $1$-d and $3$-d rectangular cavities.

In low-temperature region at $T\ll T_0$ (degenerate temperature), we see
\begin{eqnarray}\label{1d-lt-fd0}
\int_{E_0}^\infty g(E)f(E)dE&=&\int_{E_0}^\mu g(E)dE\nonumber\\
&+&\frac{\pi^2(kT)^2}{6} g'(\mu)+O((kT)^4),
\end{eqnarray}
\begin{align}\label{1d-lt-fd1}
-\int_{E_0}^\infty \varphi(E)\frac{df}{dE}dE=\varphi(\mu) +\frac{\pi^2(kT)^2}{6} \varphi''(\mu)+O((kT)^4),
\end{align}
where $g(E_0)=\varphi(E_0)=0$ is assumed.

Choosing 1-d density of states $D_1(E)$ as $g(E)$, we have
\begin{eqnarray}\label{1d-lt-num}
N&=&\int_{0}^\infty D_1(E)f(E)dE\nonumber\\
&=&\frac{\sqrt{2m}L}{\pi\hbar}\mu^{1/2}\left(1-\frac{\pi^2}{24}(kT)^2\mu^{-2}+\cdots  \right),
\end{eqnarray}
from which the chemical potential is obtained as
\begin{align}\label{1d-lt-che}
\mu=\frac{\pi^2\hbar^2}{2m} \left(\frac{N}{L}  \right)^2 \left(1+\frac{1}{3\pi^2} \left(\frac{mkT}{\hbar^2} \right)^2\left(\frac{N}{L} \right)^{-4} +\cdots \right).
\end{align}
This expansion is justified in the low-temperature and high-density regime.

In the case of the $3$-d rectangular cavity,
\begin{eqnarray}
N&=&\int_0^\infty\int_0^E dE_\parallel D_1(E_\parallel) D_2(E-E_\parallel)f(E)\nonumber\\
&=&\frac{8}{3\sqrt{2}\pi^2}\left(\frac{m}{\hbar^2} \right)^{3/2} V\mu^{3/2}\left(1+\frac{\pi^2(kT)^2}{8}\mu^{-2} \right),\nonumber\\
\end{eqnarray}
which leads to the low-temperature expansion of $\mu$ as
\begin{eqnarray}\label{3d-lt-che}
\mu&=&\frac{(3\sqrt{2})^{2/3}}{4}\pi^{4/3}\frac{\hbar^2}{m}\left(\frac{N}{V} \right)^{2/3}\nonumber\\
&\times&\left(1-\frac{4}{3} (3\sqrt{2})^{-4/3}\pi^{-2/3}\left(\frac{mkT}{\hbar^2} \right)^2 \left(\frac{V}{N}\right)^{4/3}+\cdots   \right).
\nonumber\\
\end{eqnarray}

\section{Thermodynamic averages at high-temperature region at $T\gg T_0$}\label{high-temp-myu}

In the case of high-temperature region at $T\gg T_0$,
we shall use a high-temperature expansion of Fermi-Dirac distribution
with a negative value $\mu$ as given in Eq.(\ref{fd-hy-expa}). Then
we see
\begin{align}
\int_{E_0}^\infty g(E)f(E)dE=\sum_{n=1}^\infty (-1)^{n-1}\int_{E_0}^\infty g(E) e^{-n\beta (E-\mu)}dE.
\end{align}
Choosing $1$-d density of states $D_1(E)$ as $g(E)$, we have
\begin{align}
N=\frac{L}{\sqrt{2\pi}} \sqrt{\frac{mkT}{\hbar^2}}e^{\beta\mu} \left(1-\frac{1}{\sqrt{2}}e^{\beta\mu} \right),
\end{align}
from which $\mu$ is determined by
\begin{align}\label{1d-ht-che}
e^{\beta\mu}=\sqrt{2\pi}\frac{N}{L} \frac{\hbar}{\sqrt{mkT}} \left( 1-\sqrt{\pi} \frac{  \frac{N}{L} \hbar} {\sqrt{mkT}}
+\cdots  \right).
\end{align}
This expansion is justified in the high-temperature and low-density regime.

In the case of $3$-d cavity, the particle number is
\begin{eqnarray}
N&=&\frac{2\sqrt{2}}{\pi^2}\left(\frac{m}{\hbar^2} \right)^{3/2} V \sum_{n=1}^\infty (-1)^{n-1}e^{n\beta \mu}\int_0^\infty dE E^{1/2}e^{-n\beta E}\nonumber\\
&=&\sqrt{2}\pi^{-3/2}\left(\frac{mkT}{\hbar^2} \right)^{3/2}V e^{\beta\mu}\left(1-2^{-3/2}e^{\beta\mu} \right),
\end{eqnarray}
and chemical potential is expanded as
\begin{align}\label{3d-ht-che-exp}
e^{\beta\mu}=\frac{\pi^{3/2}}{\sqrt{2}}\frac{N}{V}\left(\frac{\hbar^2}{mkT} \right)^{3/2}\left(1+\frac{\pi^{3/2}}{4}\frac{N}{V}\left(\frac{\hbar^2}{mkT}\right)^{\frac{3}{2}}+\cdots \right).
\end{align}

%


\begin{thebibliography}{99}
\bibitem{rf:carn}  S. Carnot, {\it R\'{e}flexions sur la puissance motrice du feu et sur les machines propres a
d\'{e}velopper cette puissance} (Chez Bachelier, Paris,1824).
\bibitem{rf:landau}  L. D. Landau and E. M. Lifshitz, {\it Statistical physics, third edition} (Pergamon, Oxford, 1980).
\bibitem{callen}  H. B. Callen, {\it Thermodynamics and an Introduction to Thermostatics, second edition} (John Wiley  and Sons, New York,1985).
\bibitem{rf:bend-1} C. M. Bender, D. C. Brody and B. K. Meister, J. Phys. {\bf A 33}, 4427 (2000).
\bibitem{rf:bend-2} C. M. Bender, D. C. Brody and B. K. Meister, Proc. R. Soc. Lond. {\bf A 458}, 1519 (2002).
\bibitem{rf:abe-1} S. Abe and S. Okuyama, Phys. Rev. {\bf E 83}, 021121 (2011).
\bibitem{rf:abe-2} S. Abe, Phys. Rev. {\bf E 83}, 041117 (2011).
\bibitem{rf:teif} J. Teifel and G. Mahler, Eur. Phys. J. {\bf B 75}, 275 (2010).
\bibitem{rf:quan} H. T. Quan and C. Jarzynski, Phys. Rev. {\bf E 85}, 031102 (2012).
\bibitem{rf:non-ad-f}K. Nakamura, S. K. Avazbaev, Z. A. Sobirov, D. U. Matrasulov, and T. Monnai, Phys. Rev. {\bf E 83}, 041133 (2011). This paper includes a discussion on the classical analog of the quantal non-adiabatic force.
While the classical force is based on Boltzmann's kinetic theory of an ideal gas, the quantal non-adiabatic force is due to  transitions among quantum adiabatic states, which would make unsuitable the attempt to look for the classical-quantal correspondence. By contrast the present work is purely quantum-mechanical and is not affected by the previous discussion at all.
\bibitem{rf:TUT} The technique of a time-dependent dilatation
unitary transformation for moving boundaries was first proposed in the context of heat equation theory by
R.J. Tait, Quarterly Appl. Math. {\bf 37}, 313 (1979), and was fully defined by M. Razavy, Lett. Nuovo Cimento {\bf 37}, 2384 (1983);  Phys. Rev. A {\bf 44}, 2384 (1991).
\bibitem{fric-force} The quantum non-adiabatic force has long been believed as breaking the time-reversal symmetry
and being dissipative. See for example: D.A. Hill and J.A. Wheeler, Phys. Rev. {\bf 89} 1102 (1952); M.V. Berry and J.M. Robbins,  Proc. R. Soc. London. A.  {\bf 442} 659 (1993); D. Cohen, Phys Rev Lett. {\bf 82} 4951  (1999).
\bibitem{rf:cu-ah} F. Curzon and B. Ahlborn, Am. J. Phys. {\bf 43}, 22 (1975).
\bibitem{rf:broe} C. Van den Broeck, Phys. Rev. Lett. {\bf 95}, 190602 (2005).
\bibitem{rf:sc-se}  T. Schmiedl and U. Seifert, Europhys. Lett. {\bf 81}, 20003 (2008).
\bibitem{rf:izum}  Y. Izumida and K. Okuda, Phys. Rev. {\bf E 80}, 021121 (2009).
\bibitem{rf:mas-1} S. Masuda and K. Nakamura, Proc. R. Soc. {\bf A 466}, 1135 (2010).
\bibitem{rf:mas-2} S. Masuda and K. Nakamura, Phys. Rev. {\bf A 84}, 043434 (2011).

\end{thebibliography}
\end{document}